\documentclass{article}
\usepackage{tablefootnote}
\usepackage{tabularx}
\usepackage{makecell}
\usepackage{caption}
\usepackage{graphicx}
\usepackage{amsmath, amssymb}
\usepackage{booktabs}
\usepackage{siunitx}
\usepackage{float}
\usepackage{amssymb}
\usepackage[margin=1in]{geometry}
\PassOptionsToPackage{hyphens}{url}\usepackage{hyperref}
\usepackage{xcolor}
\usepackage{tikz}
\usepackage[T1]{fontenc}

\usepackage{mathtools}

\DeclarePairedDelimiterX{\infdivx}[2]{(}{)}{%
  #1\;\delimsize\|\;#2%
}
\newcommand{\infdiv}{D_{KL}\infdivx}

\usepackage[backend=biber,
            style=alphabetic]{biblatex}
\title{Efficient Multivariate Kelly Optimization Reveals Sigmoidal Scaling Laws}
\author{Ruslan Tepelyan, Daniel Lam \\ Bloomberg \\ New York, USA\\ \texttt{rtepelyan@bloomberg.net dlam119@bloomberg.net}}
\date{}

\begin{document}

\maketitle

\begin{abstract}
For a sequence of binary bets, the Kelly criterion provides a closed-form solution that maximizes the expected growth rate of wealth. In contrast, when multiple bets are placed simultaneously (e.g., in portfolio allocation or prediction markets), the optimal Kelly strategy generally requires numerical optimization over a joint outcome space. A naive formulation scales exponentially in the number of bets, requiring $O(2^N)$ time and memory for $N$ simultaneous wagers, which restricts existing methods to small problem sizes.

We present two complementary methods that dramatically extend the scale of multivariate Kelly problems that can be solved. First, in the case of independent bets, we introduce an integral transform formulation that eliminates explicit enumeration of outcomes, reducing the computational complexity of evaluating the objective from $O(2^N)$ to $O(N)$. Combined with numerically stable quadrature, this enables accurate solutions for problems involving hundreds of bets. Second, we develop a decomposition-based approach that constructs and solves carefully chosen subproblems, yielding feasible lower bounds and infeasible upper bounds on the optimal growth rate. This provides a practical mechanism for quantifying worst-case suboptimality as a function of subproblem size.

Together, these methods make it possible to study the large-$N$ regime of the multivariate Kelly problem. Using synthetic data inspired by prediction markets, we show that the relationship between subproblem size and solution accuracy follows a simple and highly regular scaling law. In particular, the shortfall ratio between the lower and upper bounds is well-approximated by a sigmoid function of the relative subproblem size, with parameters that can be predicted from low-dimensional summary statistics of the problem.
\end{abstract}

\section{Introduction}
Given an initial endowment and a set of investment opportunities, the Kelly criterion \cite{kelly1956new} prescribes an allocation that maximizes the expected logarithm of final wealth. If investments can be repeated over successive time periods, the Kelly criterion achieves the highest possible long-run growth rate of wealth in a certain sense \cite{breiman1960investment}. These and other desirable properties have led to its successful application in a wide variety of contexts in gambling and investing \cite{thorp2008kelly}.\footnote{However, the Kelly criterion is not optimal for all investors, even in the long run. See \cite{samuelson1971fallacy,thorp1975portfolio}.} It is therefore useful to be able to compute the allocation prescribed by the Kelly criterion where this is feasible, and to obtain accurate approximations where it is not. We begin by restating the well-known case of a single risky bet.

\subsection{Single Binary Bet: Two Scenarios and Two Assets}\label{singlekelly}
Suppose there are two possible investments: a risk-free bond that we use as the numeraire asset and refer to as a ``dollar'', and a binary lottery with a payoff that depends on the outcome of a (biased) coin flip. For each dollar invested in the lottery, you win $b$ dollars with probability $p$ and lose your entire \$1 investment with probability $1-p$.\footnote{The assumption of total loss as one of the outcomes is not restrictive. If the worst loss of the risky asset is $a$ dollars per dollar invested, one can redefine the risky asset as a combination of the original risky asset and the risk-free bond with weights $1/a$ and $1-1/a$, respectively.} We assume that $p(1+b)>1$, giving a positive ``edge''. The decision variable is the fraction $f$ of wealth invested in the lottery, with the remaining $1-f$ invested in the bond. Without loss of generality, assume an initial endowment of one dollar. The expected log wealth is then
\begin{equation}\label{eq:uwealth}
U(f) = p \log(1+fb) + (1-p) \log(1-f).
\end{equation}
This is maximized at
$$
f^* = p - \frac{1-p}{b}.
$$
Under this parameterization, the optimal \textit{Kelly fraction} is always strictly less than the probability of winning.

Additional insight can be obtained by re-parameterizing the lottery in terms of its expected return $\mu$ and variance $\sigma^2$:
\begin{align*}
\mu = p (1+b) - 1, \
\sigma^2 = p (1-p) (1+b)^2.
\end{align*}
A positive edge corresponds to $\mu>0$, and the Kelly fraction becomes
\begin{equation}
f^* = \frac{\mu}{\mu + \frac{\sigma^2}{1+\mu}}.
\end{equation}
This expression makes explicit the trade-off between expected return and risk. For fixed $\sigma^2$,
\begin{align*}
f^* \to \frac{\mu}{\sigma^2} \quad \text{as } \mu \to 0, \
f^* \to 1 \quad \text{as } \mu \to \infty.
\end{align*}
The quantity $\mu/\sigma^2$ coincides with the optimal risky fraction in the continuous-time Merton problem with logarithmic utility \cite{merton1969lifetime}. Thus, Kelly betting is approximated by the Merton rule for small edges and converges to full investment in the risky asset as the edge becomes large.

\subsection{General Case: $K$ Scenarios and $N+1$ Assets}

In the general discrete setting, there are $K$ scenarios and $N$ risky assets in addition to the risk-free bond. Let the net return (gross payoff minus price paid) of asset $i$ in scenario $k$ be $R_{ki}$, and collect these in a $K \times N$ matrix $R$. The Kelly criterion prescribes a unique portfolio provided there is no arbitrage (no portfolio $w$ such that $w^T R_k \ge 0$ for all $k$, with strict inequality for at least one scenario) and the assets are non-redundant (the columns of $R$ are linearly independent). In general, however, there is no analytical solution unless the market is complete ($K=N+1$) \cite{vander1977strategy}, as in the single-bet case of Section~\ref{singlekelly}.

Let $p_k$ denote the probability of scenario $k$. The expected logarithm of final wealth for a portfolio $w$ is
$$
U(w) = \sum_{k=1}^{K} p_k \log\left(1 + w^T R_k\right)
$$
where $R_k$ denotes the $k$th row of $R$. The first-order conditions are given by
\begin{equation}\label{eq:foc}
\nabla U(w) = \sum_{k=1}^{K} \frac{p_k R_k}{1 + w^T R_k} = 0.
\end{equation}
Thus, the Kelly criterion reduces to solving a system of $N$ nonlinear equations. However, evaluating these equations requires summing over all $K$ scenarios.

We now specialize to the case of $N \geq 2$ independent binary lotteries with positive edges. The number of scenarios is $K = 2^N$, which exceeds the number of assets $N+1$, placing us in the incomplete-markets setting.

For sufficiently small $N$ (empirically, $N \lesssim 24$ on modern hardware), one can compute the optimal solution $w^*$ to \eqref{eq:foc} using direct numerical optimization. Each lottery is normalized to lose one dollar in its worst outcome, so each asset can be viewed as a limited-liability investment priced at \$1, and $w_i$ represents the fraction of wealth allocated to asset $i$. These weights are positive and must sum to less than one to avoid ruin in the worst-case scenario.

A brute-force approach enumerates all $K$ scenarios, constructing each return vector $R_k$ along with its probability $p_k$. Given $R$ and $p$, one can evaluate gradients and Hessians of $U(w)$ and apply standard optimization techniques.

In practice, the performance of such methods depends sensitively on implementation details, including initialization, line search strategy, and parameterization (e.g., direct weights versus softmax representations). These issues are exacerbated when the total Kelly fraction $\sum_i w_i^*$ is large. In such regimes, the worst-case term $1 + w^T R_k$ approaches zero, leading to ill-conditioning and numerical instability.

The fundamental limitation, however, is combinatorial: evaluating \eqref{eq:foc} requires summing over $K = 2^N$ scenarios, making direct optimization infeasible for even moderately large $N$. This motivates the development of methods that (i) avoid explicit enumeration of outcomes and (ii) provide controlled approximations with quantifiable error.

In the remainder of the paper, we introduce two such approaches. First, we develop an integral transform method that eliminates the need to sum over exponentially many scenarios, enabling the solution of large-$N$ instances. Second, we present a decomposition-based method that solves subproblems of size $n < N$ to produce lower and upper bounds on the optimal Kelly growth rate. Having established these methods, we then use them to analyze the behavior of large-scale Kelly problems, deriving empirical scaling laws under synthetic settings inspired by prediction markets.

\section{The Integral Transform Method}
\subsection{Motivation}
Our motivation in this section is to exploit the independence property of the bets in order to perform numerical optimization without any components of size $2^N$. We will achieve this via integral transforms. First, we establish some formal notational conventions.

\subsection{Problem definition and notation}

Consider $N$ independent binary bets. Bet $i$ wins with probability $p_i\in(0,1)$ and loses with probability $1-p_i$.
If you stake a fraction $w_i\ge 0$ of bankroll on bet $i$, then the \emph{net return per staked dollar} is
\[
r_i \in \{b_i,\,-1\},\qquad \mathbb{P}(r_i=b_i)=p_i,\quad \mathbb{P}(r_i=-1)=1-p_i,
\]
where $b_i>0$ is the gain on a win and you lose the full stake on a loss.

Assume all bets settle simultaneously and returns are independent across bets. The one-period wealth multiplier is
\[
X \;=\; 1 + \sum_{i=1}^N w_i r_i.
\]
This is similar to the formulation before - we have simply combined $w$ and $R$. The Kelly objective is to maximize the expected log growth
\[
f(w) \;=\; \mathbb{E}\big[\log X\big]
\]
subject to $X>0$ almost surely. A necessary condition (with only long stakes) is
\[
\sum_{i=1}^N w_i < 1.
\]

\subsection{Simplex parameterization with logits and cash}

To enforce feasibility smoothly, introduce a \emph{cash weight} $w_0$ (equivalent to our risk-free bond or numeraire from before) and parametrize all weights on the simplex:
\[
w_k(\theta) \;=\; \frac{e^{\theta_k}}{\sum_{j=0}^N e^{\theta_j}},\qquad k=0,1,\dots,N,
\]
where $w_0$ is cash (return $0$) and $w_i$ for $i\ge 1$ are bet stakes. Then $w_k\ge 0$ and $\sum_{k=0}^N w_k=1$,
so $\sum_{i=1}^N w_i = 1-w_0 < 1$ whenever $w_0>0$.

Under this simplex view, it is convenient to write the wealth multiplier as
\[
X \;=\; w_0 + \sum_{i=1}^N w_i(b_i+1)\,I_i,
\]
where $I_i\in\{0,1\}$ indicates a win: $\mathbb{P}(I_i=1)=p_i$, independent across $i$. Define
\[
c_i \;:=\; w_i(b_i+1),\qquad i=1,\dots,N,
\]
so that $X = w_0 + \sum_i c_i I_i$ and $X\ge w_0$.

\subsection{Laplace/Frullani transform and factorization}

For $x>0$, the Frullani identity for $\log$ is
\[
\log x \;=\; \int_0^\infty \frac{e^{-t}-e^{-tx}}{t}\,dt.
\]
Applying this with $x=X$ and taking expectations gives
\[
f(w) \;=\; \mathbb{E}[\log X]
\;=\; \int_0^\infty \frac{e^{-t}-\mathbb{E}[e^{-tX}]}{t}\,dt.
\]
Define the Laplace transform of $X$:
\[
Q(t) \;:=\; \mathbb{E}\!\left[e^{-tX}\right].
\]
Then
\begin{equation}\label{eq:fwint}
f(w) \;=\; \int_0^\infty \frac{e^{-t}-Q(t)}{t}\,dt.
\end{equation}

Therefore, if we could compute the Laplace transform of $X(w)$ efficiently, we would be one numerical quadrature away from computing $f(w)$.

\subsection{Stable factorization exploiting independence}
Here is the important insight: $I_i$ are \emph{independent} Bernoulli random variables. That means that for $X=w_0+\sum_i c_i I_i$, we can express the Laplace transform (a convolution of the individual bets' PMFs) as a simple product over the \emph{independent} bet distributions:

\[
Q(t) \;=\; e^{-t w_0}\prod_{i=1}^N \Big((1-p_i) + p_i e^{-t c_i}\Big).
\]
Now, define
\[
F_i(t) \;:=\; (1-p_i)+p_i e^{-t c_i},\qquad
A(t) \;:=\; \prod_{i=1}^N F_i(t),
\]
Therefore:
\begin{equation}\label{eq:qtdef}
Q(t)=e^{-t w_0}A(t)
\end{equation}

This formulation helps numerical stability. All factors satisfy $F_i(t)\in(1-p_i,1]$, hence $A(t)\in(0,1]$, and $e^{-t w_0}\in(0,1]$. This avoids numerically unstable forms that involve $e^{+t w_i}$, which will be important when we need to use quadrature techniques for computing \ref{eq:fwint}.

In practice, it is better to compute $A(t)$ in log space:
\begin{equation}\label{eq:logAt}
\log A(t)\;=\;\sum_{i=1}^N \log\!\big((1-p_i)+p_i e^{-t c_i}\big).
\end{equation}

Where the individual terms can themselves be computed via logsumexp tricks for increased numerical stability.

\subsection{Summary}
The transformation above eliminates the fundamental computational bottleneck of the multivariate Kelly problem. By exploiting the independence structure of the bets, we replace an explicit enumeration over exponentially many outcomes (a convolution) with a product representation evaluated through an integral transform.

For any given portfolio $w$, we can compute $c = w(b+1)$ directly and evaluate \eqref{eq:fwint} numerically by computing \eqref{eq:qtdef} via $\log{A(t)}$ from \eqref{eq:logAt}. Using $T$ quadrature nodes for the integration over $t$, the total computational cost is $O(TN)$. In contrast to the naive $O(2^N)$ approach, evaluation is thus linear in the number of bets, with a constant determined by the number of integration nodes required for accuracy.

At a conceptual level, this means that the exponential complexity of the original problem has been removed: we can evaluate the objective function for large-$N$ instances without enumerating scenarios. This makes it feasible to apply standard numerical optimization methods to problems of a scale that was previously inaccessible.

What remains are practical considerations required to obtain reliable and efficient solutions in this new formulation. In particular, three challenges must be addressed:
\begin{itemize}
\item Handling the numerically challenging ``almost full investment'' regime with $w_0 \ll 1$,
\item Performing the quadrature in \eqref{eq:fwint} over a large dynamic range of $t$ without requiring prohibitively large $T$ or extended precision, and
\item Efficiently evaluating not only $f(w)$, but also its gradients and Hessians for use in optimization.
\end{itemize}

We address these issues in Appendix~\ref{app:itm_details}. In the next section, we develop a complementary approach based on subproblem decomposition that provides computable bounds on the optimal solution and further extends the range of tractable problem sizes.

\section{The Subproblem Decomposition Method}
In this section, we seek to establish lower and upper bounds for the achievable expected log wealth $f(w)$ without solving the full problem. Intuitively, any feasible portfolio with positions in a subset of the available bets sets a lower bound on what can be achieved with the full set of $N$ bets. Therefore, we can solve some subproblem of size $n < N$ to get an achievable lower bound for $f(w)$. If we can make this process more informed by choosing a better set of $n$ bets, we can potentially get a better lower bound. If we can also get an upper bound, we can decide whether solving larger subproblems may be worthwhile.

We will handle the lower bound side first. We present two techniques for obtaining successive lower bounds, a stepwise optimal procedure and a looser but much faster procedure based on a greedy ranking of the bets. We will then present a method for successive upper bounds also based on bet ranking.\footnote{The lower and upper bounds are related to two different ways of representing $\log(1+x)$ as a series. The upper bound is based on numeraire transformations and is related to the well-known Taylor expansion at $x=0$, $\log(1+x) = x - \frac{x^2}{2} + \frac{x^3}{3} - ...$. The lower bound is related to the Laurent-style series decomposition $\log(1+x) = \log(x) + \log(1+\frac{1}{x}) = \log(x) + \sum\limits_{k\ge1}^{\infty}\frac{(-1)^{k+1}}{kx^k}$.}

\subsection{Stepwise Optimal Lower Bounds}
We can achieve successively tighter lower bounds on the optimal expected log wealth by constructing ``stepwise optimal portfolios'' as follows:
\begin{itemize}
    \item Consider first only portfolios that include one single lottery plus the risk-free asset; call these 1-portfolios. To find the 1-portfolio that maximizes expected log wealth, we need only to run $N$ optimizations (actually we do not need to run the numerical optimizer when there is only a single lottery, but can just use the analytical Kelly formula) and pick the one that achieves the highest expected log wealth. This gives us a lower bound.
    \item Having found the optimal 1-portfolio, we consider all portfolios obtained by adding another one of the remaining lotteries and re-optimizing the weights. To find the best such portfolio, the stepwise optimal 2-portfolio, we need to run $N-1$ numerical optimizations, each with 2 lotteries and 4 scenarios. This gives us a tighter lower bound.
    \item We can keep repeating this process. At step $n$, we find the stepwise optimal $n$-portfolio by running $N-n+1$ optimizations,\footnote{Actually it may be possible to find the stepwise optimal portfolio by running far fewer than $N-n+1$ optimizations. The optimal choice for very small and very large edges is, respectively, the lottery with the highest Sharpe ratio and the lottery with the highest expected return. We conjecture that at each step we only need to consider a lottery for possible inclusion if it is \textit{undominated} in the sense that there is no other remaining lottery with both a higher Sharpe ratio and a higher expected return.} each with $n$ lotteries and $2^n$ scenarios. We can keep this up until the number of scenarios becomes too large.
\end{itemize}

This procedure gives, after $n$ steps, not only a lower bound on the optimal expected log wealth $f(w)$, but also an explicit portfolio of $n$ lotteries that attains this bound.

\subsection{Greedy Lower Bounds}
The stepwise procedure provides very tight lower bounds, but it potentially requires a large number of solutions that grows not only as $n$ (the subproblem size we are willing to solve), but also as $N$ (the total number of bets), due to the up to $N-n+1$ different optimizations we need to check per step. Instead, we may settle for a looser bound as long as it is tight enough to be useful. We may ask when the lower bounds from the stepwise method are actually tight - it is when the overall solution is dominated by a small number of bets. That leads naturally to the greedy approach:

\begin{itemize}
    \item Rank the bets by their individual expected log wealths $U(f)$ as defined in \ref{eq:uwealth}
    \item For a given subproblem size $n$, take the top $n$ bets from the ranking and solve that subproblem
\end{itemize}

Since an optimal solution of any subset of bets $n$ is necessarily an attainable lower bound on the overall optimal solution of $N$ bets, this greedy procedure also gives an attainable lower bound. We will compare the tightness of these bounds with the ones from the stepwise optimal approach later, and show that empirically they are very close. In addition, for a given computational budget we can apply the greedy procedure to a larger subproblem immediately and get an even better lower bound that way, rather than spending the compute on evaluating the quadratically growing scenarios for the smaller subproblems in the stepwise optimal case.

\subsection{Upper Bounds}
To find an upper bound on the optimal expected log wealth, we first rank the $N$ lotteries by their expected returns $\mu$ in descending order. Then:

\begin{itemize}
    \item Replace the lottery with the highest expected return with a default-free bond having the same expected return. This is a strict improvement in the investment opportunity set, so the optimal portfolio in this modified problem gives an upper bound on the optimal expected log wealth in the original problem. The optimal portfolio in the modified problem is obviously just to hold 100\% in the now-risk-free asset, since all $N-1$ other lotteries now have nonpositive expected returns when redenominated in units of the risk-free asset.
    \item To get a tighter upper bound, we return to the original problem and replace the lottery with the second-highest expected return with a default-free bond having the same expected return. The optimal portfolio in the modified problem is just some combination of the now-risk-free asset plus the one lottery with a higher expected return. All $N-2$ other lotteries now have nonpositive expected returns when redenominated in units of the risk-free asset and can be dropped. The optimal weights are given by the analytical Kelly formula.
    \item In general, for any $n$ that is not too large, replace the lottery with the $n$-th highest expected return with a default-free bond having the same expected return. The optimal portfolio in the modified problem contains the now-risk-free asset plus the $n-1$ lotteries with a higher expected return, since the other $N-n$ lotteries all have nonpositive expected returns when redenominated in units of the risk-free asset and can be dropped. Computing this upper bound requires running only a single numerical optimization with $n-1$ lotteries and $2^{n-1}$ scenarios.
\end{itemize}

\begin{figure}[h!]
\includegraphics[width=1.0\textwidth]{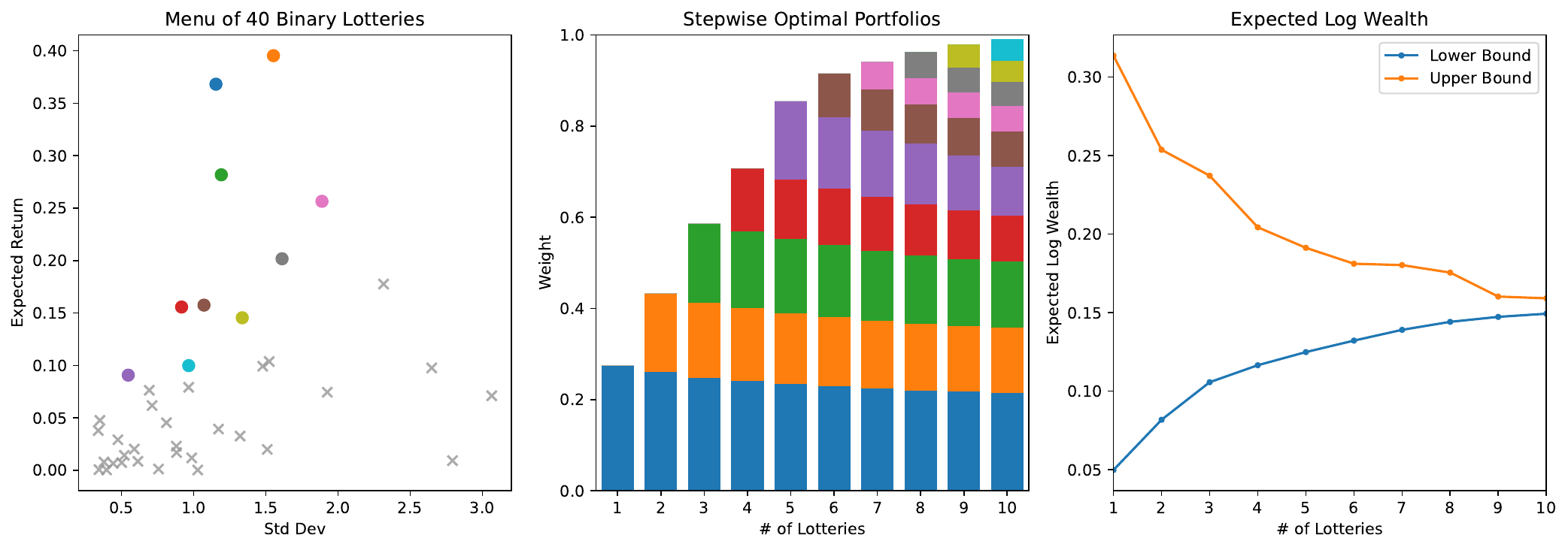}
  \caption{Illustration of the subproblem bounding approach on a representative instance with $N=40$ lotteries. Left: available lotteries in mean–standard deviation space, with selected lotteries highlighted. Middle: stepwise-optimal portfolios for increasing subproblem sizes, shown as stacked weights. Right: expected log wealth of the stepwise portfolios (lower bounds) together with corresponding upper bounds. The gap between the curves quantifies the maximum achievable improvement from including additional assets.}
  \label{fig:lower_upper_bounds_panel}
\end{figure}

In Figure~\ref{fig:lower_upper_bounds_panel}, we illustrate the behavior of the lower and upper bounds on a representative instance with $N=40$ bets and subproblem sizes up to $n=10$. 

The left panel shows the menu of available lotteries in mean–standard deviation space. The lotteries selected by the stepwise procedure are highlighted as colored circles, while the remaining lotteries are shown as $\times$ markers. 

The middle panel shows the corresponding stepwise-optimal portfolios for $n \leq 10$, with weights displayed as stacked bars. Colors are consistent with the selected lotteries in the left panel, allowing the composition of each portfolio to be tracked across steps.

The right panel plots the expected log wealth achieved by the stepwise portfolios together with the corresponding upper bounds obtained from $m$-asset optimizations for $m \leq 10$. The gap between the best lower bound and the corresponding upper bound provides a certificate of near-optimality, quantifying the maximum possible improvement from including additional lotteries beyond those already selected.

\subsection{Discussion}

The decomposition approach provides a flexible framework for constructing attainable lower bounds and unattainable upper bounds on the expected log wealth $f(w)$. Because it is agnostic to the underlying solver, it can be combined seamlessly with the integral transform method, naive numerical optimization, or other solution techniques. This makes it a useful complement to exact or approximate solvers, particularly in large-$N$ settings where computing the true optimum directly is infeasible.

In practice, these bounds can be used in several ways:

\begin{itemize}
\item \emph{Adaptive refinement.} Choose an initial subproblem size $n$ and compute the corresponding lower and upper bounds. If the gap between them is below a desired tolerance, accept the lower-bound solution; otherwise, increase $n$ according to a prescribed schedule.

\item \emph{Maximal feasible resolution.} Fix the largest subproblem size $n_{\max}$ that can be handled given computational constraints, and evaluate the corresponding bounds. The resulting gap provides a measure of worst-case suboptimality, which can inform decisions about whether additional computational resources or improved solvers are warranted.

\item \emph{Targeted accuracy.} If the scaling behavior of the \emph{shortfall ratio} $\frac{f^*_{\text{lower}}}{f^*_{\text{upper}}}$ as a function of $\frac{n}{N}$ is known or can be estimated, one can select $n$ to achieve a desired level of accuracy.

\end{itemize}

Taken together with the integral transform method, this approach provides both the ability to evaluate large-scale instances and to quantify the accuracy of approximate solutions. This combination makes it possible to study regimes that are otherwise inaccessible to direct optimization.

In the following sections, we evaluate the empirical performance of these methods and use them to investigate how the structure of optimal Kelly allocations evolves with problem size.

\section{Experimental Setup}

We design a set of synthetic experiments with two complementary objectives: (i) to validate the accuracy and numerical stability of the proposed methods on problems where ground truth is available, and (ii) to use these methods to study the scaling behavior of the multivariate Kelly solution in regimes that are otherwise computationally intractable.

To this end, our experiments address three questions:
\begin{itemize}
\item Does the integral transform method recover the optimal solution obtained via exhaustive optimization for small $N$?
\item How closely do the greedy lower bounds approximate the stepwise optimal lower bounds, and what is the resulting trade-off between accuracy and computational cost?
\item How does the subproblem size required to achieve a given accuracy scale with the number of bets and the statistical properties of disagreement?
\end{itemize}

\subsection{Synthetic Problem Parameters}

The synthetic data is inspired by prediction markets. Each bet corresponds to a single contract with:
\begin{itemize}
\item A YES contract costing $q$ and paying \$1 if the event occurs,
\item A NO contract costing $1-q$ and paying \$1 otherwise, and
\item A subjective probability $p$ representing our belief about the event.
\end{itemize}

A bet has positive value when $p \ne q$, i.e., when there is informational edge. The magnitude of this edge increases with the discrepancy between $p$ and $q$.\footnote{For a single bet, the optimal growth rate $f(w)$ is given by the Kullback–Leibler divergence $\infdiv{p}{q}$.}


To model realistic disagreement structures, we treat $q$ as an anchor and generate $p$ conditionally on $q$, reflecting the idea that large deviations from market prices are less likely than small ones. We consider two families of models:

\paragraph{Logit-offset model.}
We define
$$
\mathrm{Logit}(p) = \mathrm{Logit}(q) + \mathrm{GND}(0,\sigma^2,\beta),
$$
where $\mathrm{GND}$ is a generalized normal distribution with location $0$, scale $\sigma^2$, and shape parameter $\beta$. We consider three variance levels for the logit offset, $\mathrm{Var}(\epsilon) \in \{0.01, 0.025, 0.05\}$, corresponding to low, medium, and high disagreement. For each choice of shape parameter $\beta \in \{1,2,6\}$, we calibrate the scale parameter of the generalized normal distribution so that the variance of the offset $\epsilon$ matches these values. This calibration ensures that differences across $\beta$ reflect changes in tail behavior rather than overall scale. The shape parameters $\beta=1,2,6$ correspond to Laplace, Gaussian, and platykurtic regimes, respectively. We refer to the platykurtic regime as $GND(6)$.

\paragraph{Beta prior model.}
We alternatively draw $p \sim \mathrm{Beta}(a,b)$, with parameters chosen so that the mode equals $q$ and the mean return per bet matches that of the $low$, $medium$, and $high$ variance regimes of the logit-offset distributions. Empirically, using the mode or mean yields similar behavior; we use the mode-based parameterization in the main results.

Across both constructions, we obtain four distinct disagreement regimes at three variance levels. For each combination of regime, variance level, and problem size, we generate \SI[group-minimum-digits=4,group-separator={,},scientific-notation=false]{1000} independent instances. We sample $q \sim U(0.1, 0.9)$ to represent a broad range of market conditions away from degenerate extremes.

For validation experiments, we use $N=10$, where exhaustive optimization is feasible. For scaling experiments, we consider $N$ in increments of 20 from 20 to 200. For each instance, we compute lower and upper bounds by applying the integral transform solver to subproblems of size $n$ in increments of $\frac{N}{20}$.

\section{Validation Results}
\subsection{Confirming Solver Behavior on Small Problems}
We first verify that the integral transform method recovers the true optimal solution in regimes where exhaustive computation is feasible. To this end, we apply both the exhaustive solver and the integral transform solver to problems with $N=10$ across all disagreement regimes and variance levels.

The results are summarized in Table~\ref{tab:compsolvdiff}, which reports the maximum absolute values of several discrepancy metrics. These include the relative difference in the optimal objective value $f^*$, the difference in total leverage, the cosine difference between the optimal weight vectors $w$, and the discrepancy between the integral solver’s reported $f^*$ and an exhaustive evaluation of $f(w)$ at the same solution.

Across all regimes, these discrepancies are negligible, indicating that the integral transform method accurately recovers both the optimal value and the optimal allocation. This confirms that the transformation and numerical integration procedures introduce no material error at the scales considered.

\begin{table}[htbp]
\centering
\caption{Maximum absolute values of each difference metric across disagreement regimes and variance levels}
\label{tab:compsolvdiff}
\scriptsize
\begin{tabular}{ll S S S S}
\toprule
Disagreement regime & Variance
& {Relative in $f^*$}
& {Leverage}
& {Cosine in $w$}
& {Exh. wealth} \\
\midrule
Laplace & Low & 3.923218e-07 & 1.592709e-04 & 5.109764e-07 & 8.109832e-15 \\
Laplace & Medium & 1.349505e-07 & 1.342265e-04 & 1.251066e-07 & 9.867107e-15 \\
Laplace & High & 1.300762e-07 & 1.029047e-04 & 9.849272e-08 & 9.166279e-15 \\

Normal & Low & 3.569453e-07 & 1.680109e-04 & 3.699339e-07 & 8.906070e-15 \\
Normal & Medium & 1.711781e-07 & 2.394011e-04 & 2.628268e-07 & 1.144224e-14 \\
Normal & High & 5.730811e-08 & 1.200622e-04 & 7.016268e-08 & 8.451573e-15 \\

$GND(6)$ & Low & 2.099754e-07 & 1.341656e-04 & 1.886680e-07 & 9.605164e-15 \\
$GND(6)$ & Medium & 9.351428e-08 & 1.159903e-04 & 1.036665e-07 & 8.396062e-15 \\
$GND(6)$ & High & 7.178710e-08 & 2.105856e-04 & 2.461780e-07 & 1.228184e-14 \\

Beta & Low & 2.204286e-07 & 1.440409e-04 & 2.232665e-07 & 1.681641e-14 \\
Beta & Medium & 2.757751e-07 & 1.977230e-04 & 1.953873e-07 & 8.867906e-15 \\
Beta & High & 4.452458e-08 & 1.034206e-04 & 3.800571e-08 & 9.797718e-15 \\
\bottomrule
\end{tabular}
\end{table}

\subsection{Comparing Stepwise and Greedy Lower Bounds}
We next compare the stepwise optimal and greedy lower bound constructions on the same $N=10$ problems. While the stepwise method provides the tightest attainable lower bound, it is computationally more expensive; the greedy method is intended as a faster approximation.

Table~\ref{tab:complowdiff} reports the $90^{\text{th}}$ percentile of the relative difference in achieved $f^*$ between the two approaches for subproblem sizes $n \in {2,4,6,8}$. We focus on a high-percentile metric to capture typical worst-case behavior rather than isolated outliers.

The results show that the greedy lower bound closely tracks the stepwise optimal lower bound across all regimes and values of $n$, with discrepancies on the order of numerical precision. This indicates that the greedy construction provides a near-optimal trade-off between computational efficiency and bound tightness, making it suitable for large-scale experiments.

\begin{table}[htbp]
\centering
\caption{$90^{th}$ percentile of rel. diff. in $f^*$ between greedy and stepwise lower bounds for $N=10$}
\label{tab:complowdiff}
\scriptsize
\setlength{\tabcolsep}{3.5pt}
\begin{tabular}{ll S S S S}
\toprule
Disagreement regime & Variance & {2 bets} & {4 bets} & {6 bets} & {8 bets} \\
\midrule
Laplace & Low & 7.168294e-14 & 7.401394e-14 & 5.106783e-14 & 3.105748e-14 \\
Laplace & Medium & 1.421004e-14 & 1.552499e-14 & 1.043762e-14 & 6.370002e-15 \\
Laplace & High & 7.108540e-15 & 8.081754e-15 & 5.297796e-15 & 3.309492e-15 \\

Normal & Low & 5.016467e-14 & 6.252773e-14 & 3.991569e-14 & 2.539859e-14 \\
Normal & Medium & 1.009328e-14 & 1.259168e-14 & 8.154442e-15 & 5.308633e-15 \\
Normal & High & 5.229812e-15 & 6.803688e-15 & 4.155494e-15 & 2.775668e-15 \\

$GND(6)$ & Low & 6.072273e-14 & 5.322680e-14 & 3.385301e-14 & 2.204353e-14 \\
$GND(6)$ & Medium & 1.244061e-14 & 1.080696e-14 & 6.891134e-15 & 4.590691e-15 \\
$GND(6)$ & High & 6.363798e-15 & 5.382125e-15 & 3.597711e-15 & 2.426541e-15 \\

Beta & Low & 5.330088e-14 & 6.023236e-14 & 3.382349e-14 & 2.437616e-14 \\
Beta & Medium & 2.775961e-14 & 3.224844e-14 & 1.731082e-14 & 1.199019e-14 \\
Beta & High & 6.285996e-15 & 6.076047e-15 & 4.237038e-15 & 2.587296e-15 \\
\bottomrule
\end{tabular}
\end{table}

Taken together, these results validate both the accuracy of the integral transform method and the practical effectiveness of the greedy lower bound construction. With these tools in place, we now turn to the large-$N$ regime and investigate how the structure of optimal Kelly solutions scales with problem size.

\section{Empirical Scaling Laws for Large $N$}

\subsection{Motivation}

Having validated our solution methods, we now use them to study the behavior of multivariate Kelly problems in the large-$N$ regime. In particular, for problems with $N$ bets, we construct lower- and upper-bound solutions using subproblems of size $n < N$.

Our goal is to characterize the shortfall ratio
$y = \frac{f^*_{\text{lower}}}{f^*_{\text{upper}}}$
as a function of the relative subproblem size $x = \frac{n}{N}$.

A predictive model of this relationship allows us to determine the subproblem size required to achieve a desired level of accuracy.

\subsection{Functional Form and Empirical Structure}

Both $x$ and $y$ lie in $[0,1]$, with $y=0$ when $x=0$ and $y=1$ when $x=1$. This suggests a smooth, monotonic relationship. Empirically, we find that the following functional form provides an excellent fit:
\[
y \approx \mathrm{Sigmoid}(\mathrm{Logit}(x); Q, v, B),
\]
where $\mathrm{Logit}(x) = \log{\left(\frac{x}{1-x}\right)}$ and
\begin{equation}\label{eq:gensigmoid}
\mathrm{Sigmoid}(z;Q,v,B) = \left(1 + Q e^{-Bz}\right)^{-1/v}.
\end{equation}
We first examine the empirical structure of this relationship across all regimes.

\begin{figure}[htbp]
\centering
\includegraphics[width=0.75\linewidth]{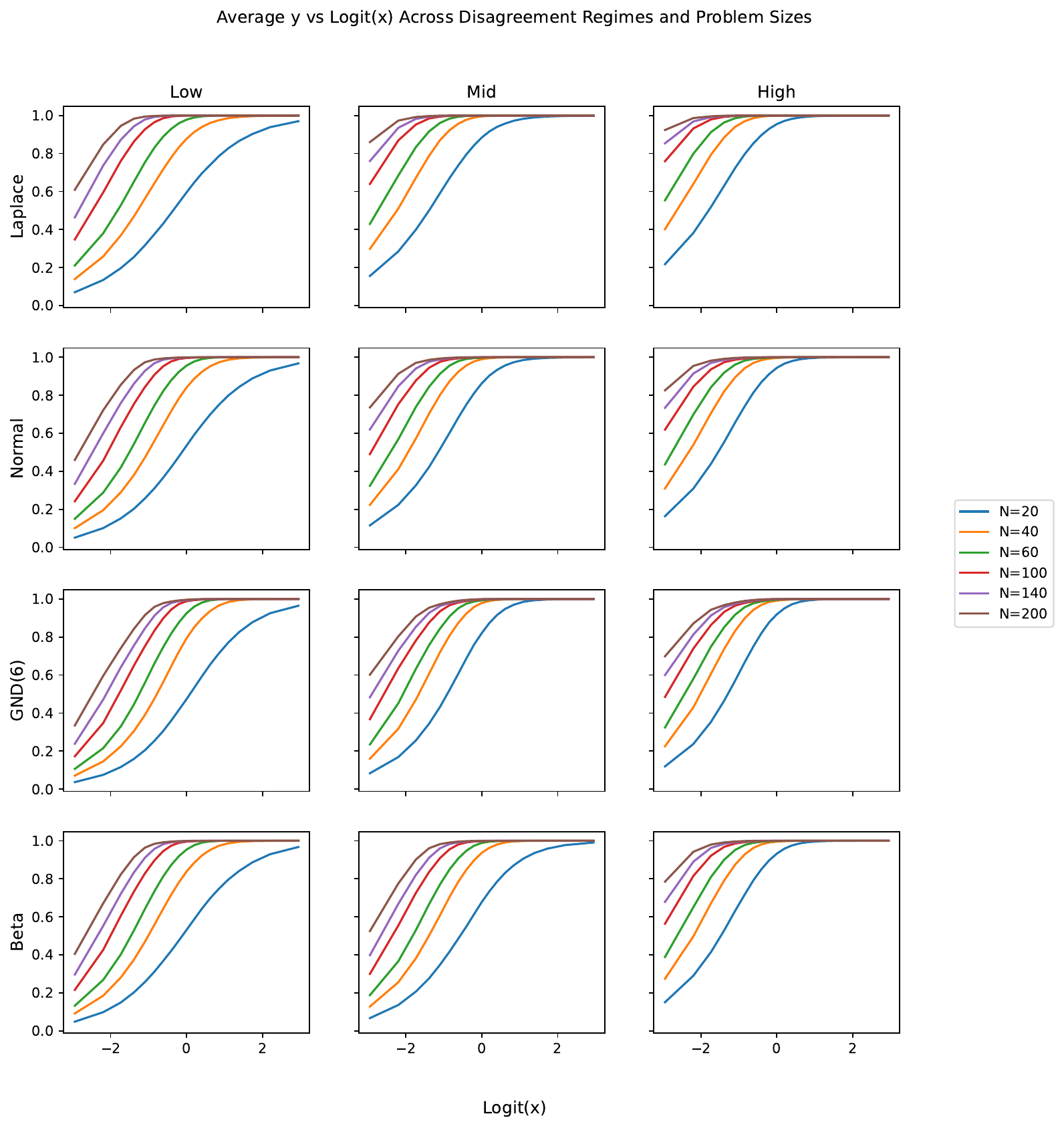}
\caption{Empirical scaling of the shortfall ratio as a function of subproblem fraction across disagreement regimes, variance levels, and problem sizes. In all cases, the relationship exhibits a consistent sigmoid structure in $\mathrm{Logit}(x)$.}
\label{fig:overall_scaling_raw}
\end{figure}

Figure~\ref{fig:overall_scaling_raw} shows that across all disagreement regimes, variance levels, and problem sizes, the scaling curves exhibit a consistent sigmoidal shape when expressed in terms of $\mathrm{Logit}(x)$. While the location and steepness vary systematically, the overall form is remarkably stable, suggesting that the behavior can be captured by a low-dimensional parametric family.

To improve numerical stability in large-$N$ settings, we model $\log{y}$ rather than $y$, since both $f^*_{\text{lower}}$ and $f^*_{\text{upper}}$ approach $1$ and direct modeling becomes sensitive to small absolute errors.

\subsection{Fit Quality and Misspecification Error}
To isolate misspecification error from parameter prediction error, we first fit the sigmoid functional form independently to each problem instance. Specifically, we optimize the parameters $(Q,v,B)$ using L-BFGS to minimize the mean squared error between $\log{y}$ and $\log{\hat{y}}$. This produces a best-fit, smooth approximation of each scaling curve within the chosen functional family, which we use as a target when modeling the dependence of $(Q,v,B)$ on problem parameters.

\begin{figure}[htbp]
\centering
\includegraphics[width=0.6\linewidth]{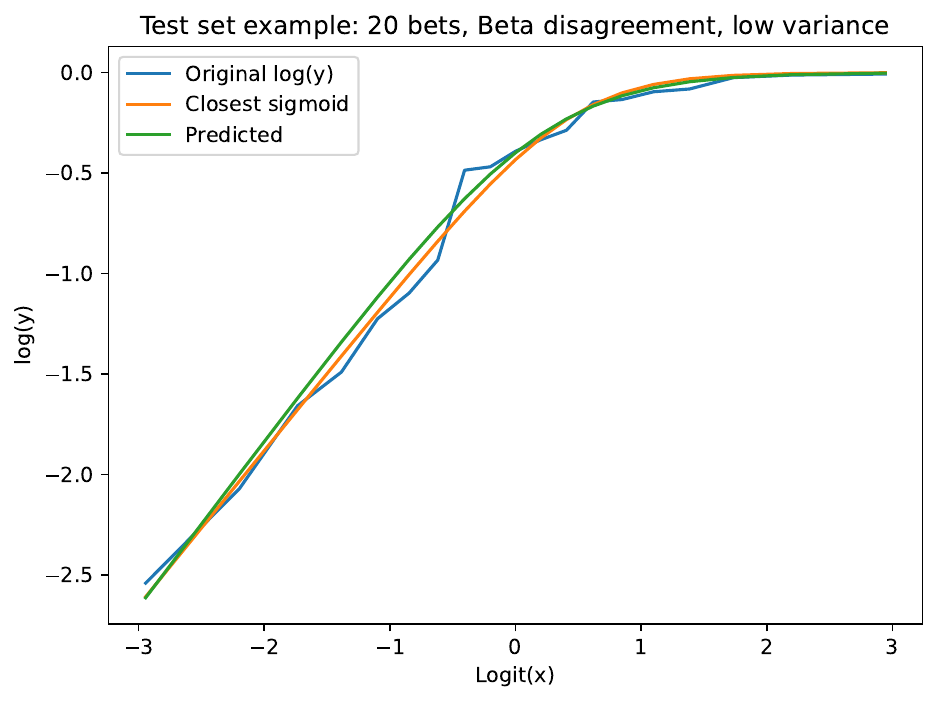}
\caption{Representative test instance showing observed $\log{y}$, best-fit sigmoid, and predicted sigmoid. The predicted curve is generated using the parameter model described in the following section. The sigmoid form closely matches the observed data, and the predicted parameters provide a strong approximation.}
\label{fig:single_example}
\end{figure}

Figure~\ref{fig:single_example} shows a representative example. The best-fit sigmoid closely matches the observed data, confirming that the functional form accurately captures the underlying scaling relationship. The predicted sigmoid also provides a strong approximation, indicating that the parameters $(Q,v,B)$ can be effectively modeled from problem features.

Across all problem instances, the misspecification error is very small. Table~\ref{tab:bestfitsigerr} summarizes these results, with median $R^2$ values exceeding $0.99$ across all regimes, confirming that the sigmoid form is an accurate description of the scaling behavior.

\begin{table}[htbp]
\centering
\caption{Scaling law misspecification error across all problems}
\label{tab:bestfitsigerr}
\scriptsize
\setlength{\tabcolsep}{3.5pt}
\begin{tabular}{S S S}
\toprule
{Mean MSE} & {Median MAE} & {1 - Median $R^2$} \\
\midrule
4.526986e-4 & 3.949906e-3 & 0.000120206 \\
\bottomrule
\end{tabular}
\end{table}

\subsection{Parameter Modeling}
Having established that the sigmoid form accurately captures the scaling relationship, we next model the parameters $(Q,v,B)$ as functions of problem characteristics.

We use a linear model with 14 input features, including $N$, $\log{N}$, and summary statistics of the differences between $p$ and $q$ and between $\mathrm{Logit}(p)$ and $\mathrm{Logit}(q)$ (mean, variance, log variance, skewness, kurtosis, and log kurtosis). Features are normalized within each training set, as defined in the next section.

To enforce positivity constraints, we model $\log{Q}$, $v'$, and $B'$ and set $v = \mathrm{softplus}(v')$ and $B = \mathrm{softplus}(B')$. The linear form is
$$
\log{Q} = \alpha_Q + X \beta_Q,
$$
with analogous expressions for $v'$ and $B'$. Parameters are estimated by minimizing mean squared error over $\log{y}$, with an $L_2$ penalty applied to the coefficient matrices $\beta_{Q,v',B'}$ but not to the intercepts $\alpha_{Q,v',B'}$.

\subsection{Training, Validation, and Test Data}

For each random seed (five in total), we split each problem type (defined by disagreement regime, variance level, and problem size) into 60/25/15 training, validation, and test sets. These splits are then aggregated across problem types. The validation set is used to select the optimal $L_2$ penalty, and test performance is averaged across seeds to obtain robust estimates of out-of-sample accuracy.

\subsection{Test Set Performance}

\begin{figure}[htbp]
\centering
\includegraphics[width=0.6\linewidth]{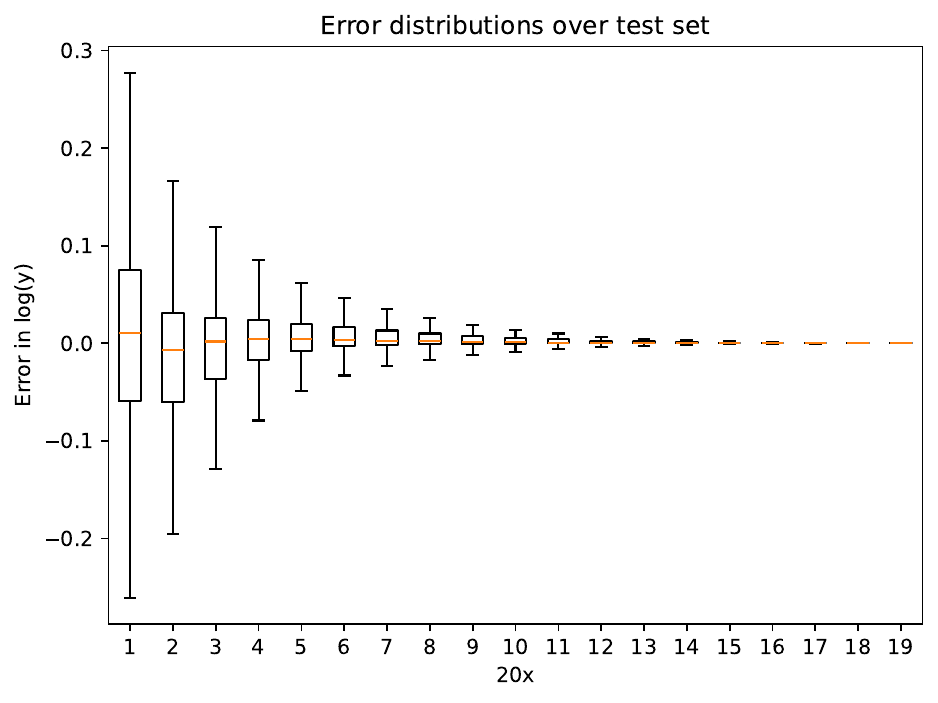}
\caption{Distribution of prediction errors in $\log{y}$ across subproblem fractions. Errors decrease and concentrate as $x=n/N$ increases.}
\label{fig:error_boxes}
\end{figure}

Figure~\ref{fig:error_boxes} summarizes prediction errors across all test instances. The error decreases and becomes more tightly concentrated as $x$ increases, reflecting the increasing regularity of the scaling behavior in the large-$N$ regime.

Tables~\ref{tab:test_perf_regimes} and~\ref{tab:test_perf_numbets} provide detailed performance metrics across regimes and problem sizes. The model achieves consistently high accuracy, with median $R^2$ values of approximately $0.98$ across all settings.

\begin{table}[htbp]
\centering
\caption{Average performance of linear model on test set regimes}
\label{tab:test_perf_regimes}
\scriptsize
\begin{tabular}{ll S S S S}
\toprule
Disagreement regime & Variance
& {Mean MSE}
& {Median MAE}
& {Median $R^2$} \\
\midrule
Laplace & Low & 0.005011 & 0.019918 & 0.976071 \\
Laplace & Medium & 0.002561 & 0.009017 & 0.954618 \\
Laplace & High & 0.001881 & 0.006314 & 0.931909 \\

Normal & Low & 0.004395 & 0.019399 & 0.988576 \\
Normal & Medium & 0.002432 & 0.010493 & 0.977426 \\
Normal & High & 0.001855 & 0.008260 & 0.966286 \\

$GND(6)$ & Low & 0.004157 & 0.018670 & 0.993720 \\
$GND(6)$ & Medium & 0.002717 & 0.014620 & 0.981781 \\
$GND(6)$ & High & 0.002190 & 0.012426 & 0.972644 \\

Beta & Low & 0.005172 & 0.018911 & 0.989260 \\
Beta & Medium & 0.004115 & 0.015470 & 0.985716 \\
Beta & High & 0.002226 & 0.008443 & 0.971245 \\
\midrule
All & All & 3.226e-03 & 1.338e-02 & 0.97932 \\
\bottomrule
\end{tabular}
\end{table}

\begin{table}[htbp]
\centering
\caption{Average performance of linear model on test set problem sizes}
\label{tab:test_perf_numbets}
\scriptsize
\begin{tabular}{l S S S S}
\toprule
{Number of bets $N$}
& {Mean MSE}
& {Median MAE}
& {Median $R^2$} \\
\midrule
20 & 0.017716 & 0.074446 & 0.973034 \\
40 & 0.005739 & 0.035830 & 0.981677 \\
60 & 0.002761 & 0.021973 & 0.983664 \\
80 & 0.001765 & 0.016332 & 0.983034 \\
100 & 0.001193 & 0.012915 & 0.982697 \\
120 & 0.000884 & 0.010520 & 0.981291 \\
140 & 0.000686 & 0.009038 & 0.978663 \\
160 & 0.000593 & 0.008094 & 0.976773 \\
180 & 0.000492 & 0.007368 & 0.974975 \\
200 & 0.000432 & 0.006642 & 0.972296 \\
\midrule
All & 3.226e-03 & 1.338e-02 & 0.97932 \\
\bottomrule
\end{tabular}
\end{table}

\subsection{Collapse to a Universal Scaling Law}

The strongest evidence for a common underlying structure is obtained by examining the scaling curves after reparameterization.

\begin{figure}[htbp]
\centering
\includegraphics[width=0.7\linewidth]{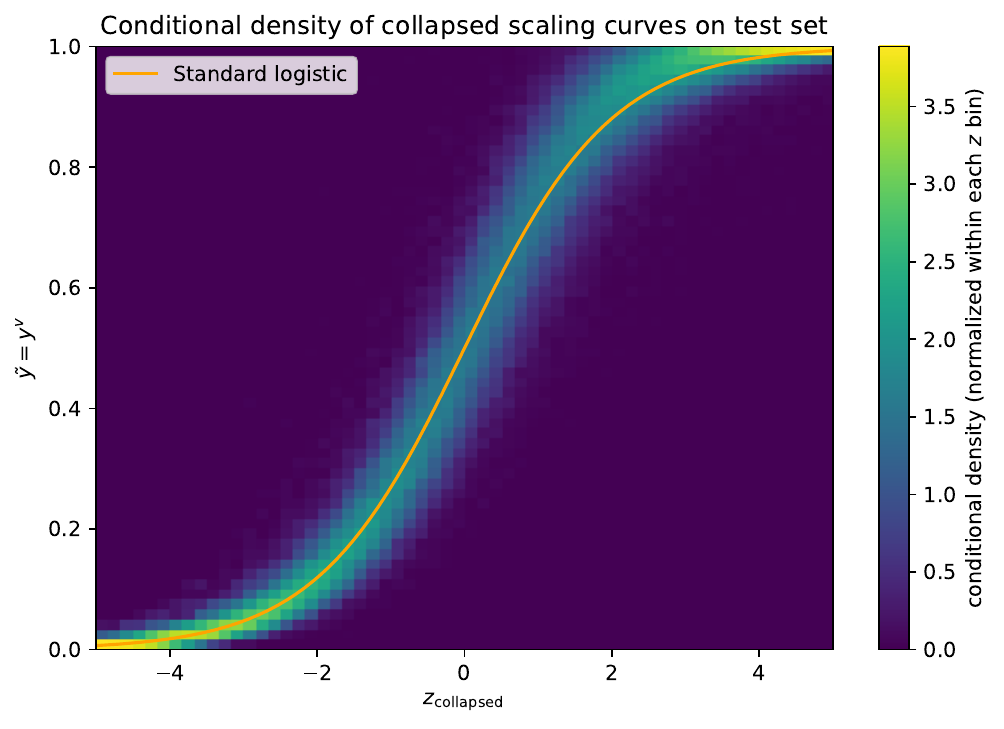}
\caption{Conditional density of transformed scaling curves after reparameterization. Each vertical slice is normalized to represent the distribution of $\tilde y = y^v$ given the collapsed coordinate. The data collapses onto the standard logistic curve. Color intensity reflects a monotone transformation of the conditional density to enhance visual contrast.}
\label{fig:collapse}
\end{figure}

Figure~\ref{fig:collapse} shows the conditional distribution of the transformed response $\tilde y = y^v$ as a function of the collapsed coordinate
\[
z_{\mathrm{collapsed}} = B\,\mathrm{Logit}(x) - \log Q.
\]

After this transformation, scaling curves from all problem instances align closely with the standard logistic function of $z_{collapsed}$ via substitution in \ref{eq:gensigmoid}. The conditional density is tightly concentrated around the standard curve across the full range of $z_{\mathrm{collapsed}}$, with no significant systematic deviations.

This demonstrates that the variability across regimes, variance levels, and problem sizes is largely captured by the parameters $(Q,v,B)$, and that the underlying scaling behavior is governed by a common low-dimensional structure.

\subsection{Discussion}

Taken together, these results show that the trade-off between subproblem size and solution accuracy in multivariate Kelly problems is both highly regular and predictable. The scaling behavior is well-approximated by a sigmoid function, its parameters can be accurately modeled using simple summary statistics, and, after reparameterization, the resulting curves collapse onto a universal form.

This structure enables practical selection of subproblem sizes to achieve desired accuracy levels in large-scale settings, and suggests that seemingly complex multivariate Kelly problems exhibit a surprising degree of underlying regularity.

\section{Conclusion and Future Work}
We have presented two complementary approaches for solving the multivariate Kelly problem and used them to analyze its behavior at scales that are inaccessible to naive methods. First, we introduced an integral transform formulation that eliminates the need to enumerate exponentially many outcomes, reducing the computational complexity from $O(2^N)$ to $O(N)$ per function evaluation. With appropriate numerical stabilization, this method enables accurate solutions for problems involving hundreds of bets. Second, we developed a decomposition-based approach that constructs tractable subproblems and provides computable lower and upper bounds on the optimal growth rate, allowing practitioners to quantify solution quality as a function of computational effort.

Together, these methods make it possible to study the large-$N$ regime of the multivariate Kelly problem. Using synthetic data inspired by prediction markets, we showed that the relationship between subproblem size and solution accuracy follows a simple and highly regular scaling law. In particular, the shortfall ratio is well-approximated by a sigmoid function of the relative subproblem size, with parameters that can be predicted from low-dimensional summary statistics of the problem.

Several directions for future work remain. An important extension is to relax the assumption of independence between bets and develop analogous methods for correlated settings. It would also be of interest to investigate whether similar scaling laws hold under broader classes of disagreement distributions, and to better understand the theoretical origins of the observed sigmoid structure. Optimal solutions for large $N$ tend to be nearly saturated (with leverage close to 1); it would be interesting to explore whether constraining the leverage has a large impact on the relative weights. Finally, it would be interesting to compare large-$N$ solutions from this method to various heuristics and to machine learning models trained on optimal solutions.

Overall, these results suggest that large-scale multivariate Kelly problems, while combinatorial in formulation, exhibit regular and predictable structure when viewed through the lens of the methods developed here.

\printbibliography[title={References}]

\clearpage
\appendix

\section{Optimizing via the Integral Transform Method}\label{app:itm_details}
 
\subsection{Tail subtraction for nearly full investment}
The challenging regime (from an optimization standpoint) is $w_0\approx 0$, where the integrals have a long tail on scale $t\sim 1/w_0$. Here, we present a way to mitigate the long tail arising from $w_0\approx 0$.

Let
\[
q_0 \;:=\; \mathbb{P}(I_1=\cdots=I_N=0) \;=\; \prod_{i=1}^N (1-p_i).
\]
The \emph{all-loss} outcome has $X=w_0$ with probability $q_0$. As $t\to\infty$,
\[
Q(t)\sim q_0 e^{-t w_0},
\]
which produces slow-decaying tails in $\int (e^{-t}-Q(t))/t\,dt$ and in derivatives.

Define a reference transform matching both $t\to 0$ and the $t\to\infty$ tail:
\[
Q_{\mathrm{ref}}(t) \;:=\; (1-q_0)e^{-t} + q_0 e^{-t w_0}.
\]
Then $Q_{\mathrm{ref}}(0)=1=Q(0)$ and $Q_{\mathrm{ref}}(t)\sim q_0 e^{-t w_0}$ as $t\to\infty$.
Using Frullani's identity,
\[
\int_0^\infty \frac{e^{-t}-Q_{\mathrm{ref}}(t)}{t}\,dt
\;=\; q_0\int_0^\infty \frac{e^{-t}-e^{-t w_0}}{t}\,dt
\;=\; q_0 \log w_0.
\]
Therefore,
\[
\boxed{
f(w) \;=\; q_0 \log w_0 \;+\; \int_0^\infty \frac{Q_{\mathrm{ref}}(t)-Q(t)}{t}\,dt.
}
\]
Since $Q_{\mathrm{ref}}(t)-Q(t)$ cancels the slowest tail, the remainder decays on the scale
$t\sim 1/(w_0+c_{\min})$ where $c_{\min}=\min_i c_i$ among non-negligible positions.

Using $Q(t)=e^{-t w_0}A(t)$,
\[
Q_{\mathrm{ref}}(t)-Q(t)
\;=\; (1-q_0)e^{-t} + e^{-t w_0}\big(q_0 - A(t)\big).
\]

We can now already efficiently compute $f(w)$. But for challenging problems, we will require additional steps to ensure that the numerical quadratures used will converge stably. Even with tail subtraction, we may still need a large dynamic range for the integration variable $t$.

\subsection{Double-exponential quadrature}

To evaluate integrals $\int_0^\infty g(t)\,dt$ efficiently over a wide dynamic range like on $[0,\infty)$, we use the
double-exponential (DE) change of variables
\[
t \;=\; \exp(\sinh u),\qquad
\frac{dt}{du} \;=\; \cosh u \,\exp(\sinh u) \;=\; \cosh u\, t.
\]
Then
\[
\int_0^\infty g(t)\,dt \;=\; \int_{-\infty}^{\infty} g(\exp(\sinh u))\,\cosh u\,\exp(\sinh u)\,du.
\]
We can approximate the $u$-integral by the trapezoid rule on $u_k = kh$, $k=-M,\dots,M$:
\[
\int_0^\infty g(t)\,dt
\;\approx\;
\sum_{k=-M}^M W_k\, g(t_k),
\quad
t_k=\exp(\sinh u_k),
\quad
W_k = h\,\omega_k\,\cosh u_k\, t_k,
\]
where $\omega_k=1$ except $\omega_{\pm M}=\tfrac12$ for trapezoid endpoints.

As $|u|\to\infty$, $t=\exp(\sinh u)$ grows (or shrinks) double-exponentially, so the grid covers both
small and extremely large $t$ with moderate $M$.

Now we can put everything together to evaluate $f(w)$ quickly, efficiently, and correctly.

\subsection{Evaluating $f$}

Using tail subtraction,
\[
\boxed{
f(w) \;=\; q_0 \log w_0 \;+\;
\int_0^\infty
\frac{(1-q_0)e^{-t} + e^{-t w_0}(q_0-A(t))}{t}\,dt.
}
\]
This is evaluated by DE quadrature with
\[
g_{\mathrm{obj}}(t) \;=\; \frac{(1-q_0)e^{-t} + e^{-t w_0}(q_0-A(t))}{t}.
\]

We can apply similar techniques to get the gradient and even Hessian of the problem, allowing us to apply standard optimization techniques.

\subsection{Gradient with respect to bet weights}

First note
\[
\frac{\partial f}{\partial w_i}
\;=\;
\mathbb{E}\!\left[\frac{r_i}{X}\right],
\qquad
r_i = -1 + (b_i+1) I_i.
\]
Use the Laplace identity for $1/X$:
\[
\frac{1}{X}=\int_0^\infty e^{-tX}\,dt,
\]
so
\[
\frac{\partial f}{\partial w_i}
=
\int_0^\infty \mathbb{E}\!\left[r_i e^{-tX}\right]\,dt.
\]
With $z_i(t):=e^{-t c_i}$ and $F_i(t)=(1-p_i)+p_i z_i(t)$, define
\[
a_i(t) \;:=\; \frac{\mathbb{E}[r_i e^{-t c_i I_i}]}{\mathbb{E}[e^{-t c_i I_i}]}
=
\frac{p_i b_i z_i(t) - (1-p_i)}{F_i(t)}
\;=\; -1 + u_i(t),
\]
where the stable nonnegative quantity
\[
\boxed{
u_i(t) \;:=\; \frac{p_i (b_i+1)\, z_i(t)}{F_i(t)} \;\ge 0.
}
\]
Then
\[
\boxed{
g_i := \frac{\partial f}{\partial w_i}
\;=\;
\int_0^\infty e^{-t w_0}A(t)\,a_i(t)\,dt.
}
\]

\subsubsection{Gradient tail subtraction}
Similar to what we did for tail subtraction in $f(w)$, we can improve the numerical behavior for the gradient.

As $t\to\infty$, $A(t)\to q_0$ and $a_i(t)\to -1$, hence
\[
g_i \sim \int_0^\infty \big(-q_0 e^{-t w_0}\big)\,dt = -\frac{q_0}{w_0}.
\]
Define the remainder $g_i^{\mathrm{rem}}$ by
\[
\boxed{
g_i
\;=\;
-\frac{q_0}{w_0}
\;+\;
g_i^{\mathrm{rem}},
\qquad
g_i^{\mathrm{rem}}
=
\int_0^\infty e^{-t w_0}\Big((q_0-A(t)) + A(t)u_i(t)\Big)\,dt.
}
\]
The integrand in $g_i^{\mathrm{rem}}$ decays on the faster scale $1/(w_0+c_{\min})$.

\subsubsection{Gradient with respect to logits $\theta$}
In practice, working with logits usually leads to better convergence behavior. Let $w(\theta)$ be the softmax weights on $\{0,\dots,N\}$ and define the softmax Jacobian
\[
J(\theta) \;=\; \nabla_\theta w
\;=\;
\mathrm{diag}(w) - w w^\top.
\]
Cash has return $0$, so $\partial f/\partial w_0 = 0$. Define an augmented gradient
$\tilde g \in \mathbb{R}^{N+1}$ by
\[
\tilde g_0 = 0,\qquad \tilde g_i = g_i,\; i=1,\dots,N.
\]
Then
\[
\boxed{
\nabla_\theta f(\theta) \;=\; J(\theta)\,\tilde g.
}
\]
A numerically stable implementation avoids explicitly forming the large $-q_0/w_0$ term by using
$g^{\mathrm{rem}}$ and the identity above.

Let $w_0$ be cash, $w\in\mathbb{R}^N$ be bet weights, and define
\[
\hat m \;:=\; \sum_{i=1}^N w_i\, g_i^{\mathrm{rem}}.
\]
Then the logit-gradient components can be written stably as
\[
\boxed{
\frac{\partial f}{\partial \theta_i}
=
w_i\Big(g_i^{\mathrm{rem}} - (\hat m + q_0)\Big),\qquad i=1,\dots,N,
}
\]
\[
\boxed{
\frac{\partial f}{\partial \theta_0}
=
q_0(1-w_0) - w_0 \hat m.
}
\]
These formulas are algebraically equivalent to $J\tilde g$ but avoid catastrophic cancellation
when $w_0\ll 1$.

\subsection{Hessian structure and Newton--CG in $\theta$-space}

\subsubsection{Hessian in $w$-space}

The $w$-space Hessian is
\[
\boxed{
H^{(w)}_{ij} \;=\; \frac{\partial^2 f}{\partial w_i \partial w_j}
\;=\;
-\mathbb{E}\!\left[\frac{r_i r_j}{X^2}\right],
}
\]
so $H^{(w)}$ is negative semidefinite (concavity of $\mathbb{E}[\log X]$).

Using $1/X^2 = \int_0^\infty t e^{-tX}\,dt$,
\[
H^{(w)}_{ij} = -\int_0^\infty t\,\mathbb{E}[r_i r_j e^{-tX}]\,dt.
\]

Define additionally
\[
m_i(t) \;:=\; \frac{\mathbb{E}[r_i^2 e^{-t c_i I_i}]}{\mathbb{E}[e^{-t c_i I_i}]}
=
\frac{(1-p_i)+p_i b_i^2 z_i(t)}{F_i(t)},
\]
and
\[
d_i(t) \;:=\; m_i(t) - a_i(t)^2
=
\frac{p_i(1-p_i)\,z_i(t)\,(b_i+1)^2}{F_i(t)^2}\;\ge 0.
\]
Then at each $t$ the integrand matrix has the form
\[
\boxed{
H^{(w)}(t)
=
-t\,e^{-t w_0}A(t)\Big(a(t)a(t)^\top + \mathrm{diag}(d(t))\Big),
}
\]
and $H^{(w)} = \int_0^\infty H^{(w)}(t)\,dt$.

\subsubsection{Hessian--vector product in $w$-space}

For any $v\in\mathbb{R}^N$,
\[
\boxed{
H^{(w)}v
=
-\int_0^\infty t\,e^{-t w_0}A(t)\Big(a(t)\,(a(t)^\top v) + d(t)\odot v\Big)\,dt.
}
\]
This can be evaluated by DE quadrature in $O(NJ)$ per HVP (where $J$ is the number of DE nodes),
without forming the full dense Hessian.

\subsubsection{Hessian tail structure and stable subtraction}

As $t\to\infty$, $A(t)\to q_0$, $a(t)\to -\mathbf{1}$, and $d(t)\to 0$, so the slow tail corresponds to
\[
H^{(w)}_{\text{tail}}(t)\approx -t\,q_0\,e^{-t w_0}\,\mathbf{1}\mathbf{1}^\top,
\quad\Rightarrow\quad
H^{(w)}_{\text{tail}} \approx -\frac{q_0}{w_0^2}\mathbf{1}\mathbf{1}^\top.
\]
Thus
\[
H^{(w)}_{\text{tail}}v \;=\; -\frac{q_0}{w_0^2}\,(\mathbf{1}^\top v)\,\mathbf{1},
\]
a rank-1 term that can be handled analytically. The remainder integral, written in terms of
$u(t)=a(t)+\mathbf{1}$ and $q_0-A(t)$, decays on the faster scale $1/(w_0+c_{\min})$.

\subsubsection{Hessian in $\theta$-space and Hessian--vector products}

Let $\tilde g\in\mathbb{R}^{N+1}$ be the augmented $w$-gradient ($\tilde g_0=0$, $\tilde g_{1:N}=g$).
By chain rule,
\[
H^{(\theta)}
=
\nabla_\theta^2 f(\theta)
=
J\,\tilde H^{(w)}\,J \;+\; \sum_{k=0}^N \tilde g_k \,\nabla_\theta^2 w_k,
\]
where $\tilde H^{(w)}$ is the $(N+1)\times(N+1)$ matrix with $0$ rows/cols for the cash dimension and
$H^{(w)}$ in the bet block.

In practice one does not form $H^{(\theta)}$; instead use Hessian--vector products (HVPs) for Newton--CG.
For any $v\in\mathbb{R}^{N+1}$ define $\delta w = Jv$ (so $\sum_k \delta w_k=0$). Then an HVP can be written as
\[
\boxed{
H^{(\theta)}v
=
J\,[0;\,H^{(w)}\delta w_{1:N}]
\;+\;
(dJ(v))\,\tilde g,
}
\]
where $(dJ(v))$ is the directional derivative of $J$ in direction $v$. A convenient closed form is
\[
\boxed{
(dJ(v))\,\tilde g
=
\delta w \odot \tilde g
\;-\;
\delta w\, (w^\top \tilde g)
\;-\;
w\,(\delta w^\top \tilde g),
}
\]
with all quantities in $\mathbb{R}^{N+1}$ and $\odot$ elementwise multiplication.
In the small-$w_0$ regime, evaluate $\delta w^\top \tilde g$ using the tail-subtracted $g^{\mathrm{rem}}$ to avoid
subtracting large nearly equal terms.

\subsubsection{Newton--CG with damping}

Because $f$ is concave in $w$ (and typically in $\theta$ modulo gauge), the Newton system for maximizing $f$ is
\[
(-H^{(\theta)} + \lambda I)\,\Delta\theta \;=\; \nabla_\theta f,
\]
where $\lambda>0$ is a Levenberg/LM damping term to ensure the operator is positive definite on the relevant subspace.
Conjugate gradients (CG) solves this system using only HVPs $v\mapsto H^{(\theta)}v$.

A line search (Armijo or Wolfe) on $\theta \leftarrow \theta + \alpha \Delta\theta$ is used to guarantee ascent.
Because softmax is invariant to adding a constant to all logits, it is convenient to fix the gauge by projecting
$\theta \leftarrow \theta - \frac{1}{N+1}\mathbf{1}^\top\theta\,\mathbf{1}$ after each update.

\subsection{Summary}
In this Appendix, we have provided important details for how to get the computation of the various components of the Integral Transform Method to actually converge and work in practice. Here is a summary of the overall method components:

\begin{itemize}
\item Independence is exploited via the Laplace transform $Q(t)=e^{-t w_0}\prod_i[(1-p_i)+p_i e^{-t c_i}]$.
\item Frullani's identity turns $\mathbb{E}[\log X]$ into a one-dimensional integral.
\item When $w_0$ is tiny, explicit tail subtraction using $q_0=\prod_i(1-p_i)$ eliminates the slow $e^{-w_0 t}$ tail from
      the numerical remainder.
\item Double-exponential quadrature on $t\in[0,\infty)$ provides rapid convergence across the large dynamic range in $t$.
\item Gradients and Hessian information are computed in stable, tail-subtracted forms; Newton--CG in $\theta$-space uses HVPs
      and avoids forming dense Hessians.
\end{itemize}

\end{document}